\documentclass[sigconf]{acmart}

\AtBeginDocument{%
  }

\usepackage[subtle]{savetrees}
\usepackage{xcolor}
\usepackage{svg}

\usepackage{xspace}
\usepackage{amsmath}
\usepackage{amsthm}
\usepackage{amstext}
\usepackage{amsfonts}
\usepackage{tikz}
\usepackage{subcaption}

\usepackage{algorithm}
\usepackage{algpseudocode}
\usepackage{wrapfig}
\usepackage{url}
\usepackage{enumerate}
\usepackage{enumitem}
\usepackage[normalem]{ulem}
\usepackage{relate}
\usepackage{marginnote}
\usepackage{balance}
\usepackage[off]{our-comments}

\renewcommand{\epsilon}{\varepsilon}

\newcommand{\defn}[1]		{{\textit{\textbf{\boldmath #1}}}\xspace}

\addCommenter{prashant}{Prashant}{PP}{red}
\addCommenter{rob}{Rob}{RJ}{violet}
\addCommenter{michael}{Michael}{MB}{blue}
\addCommenter{Michael}{Michael}{MB}{blue}
\addCommenter{mab}{Michael}{MB}{blue}
\addCommenter{martin}{Martin}{MFC}{green}
\addCommenter{alex}{Alex}{AC}{orange}

\definecolor{magenta4}{rgb}{0.5625,0,0.5625}
\definecolor{green4}{rgb}{0,0.5625,0}
\definecolor{orange4}{rgb}{0.98,0.31,0.09}

\newcommand{\depricate}[1]{}

\newcommand{\para}[1]{\smallskip\noindent\textbf{#1}}

\newcommand{\calS}{\mathcal{S}}

\makeatletter
\algrenewcommand\ALG@beginalgorithmic{\footnotesize}
\makeatother

\algnewcommand{\algorithmicor}{\textbf{ or }}
\algnewcommand{\OR}{\algorithmicor}

\newcommand{\kmer}{$k$-mer\xspace}

\newcommand{\kmers}{$k$-mers\xspace}

\begin{document}


\title{Time To Replace Your Filter: How Maplets Simplify System Design}


\author{Michael A. Bender}
\affiliation{
  \institution{Stony Brook University}
  \country{USA}
}
\email{bender@cs.stonybrook.edu}

\author{Alex Conway}
\affiliation{
  \institution{Cornell Tech}
  \country{USA}
}
\email{me@ajhconway.com}

\author{Mart\'{\i}n Farach-Colton}
\affiliation{
  \institution{New York University}
  \country{USA}
}
\email{martin@farach-colton.com}

\author{Rob Johnson}
\affiliation{
  \institution{VMware Research}
  \country{USA}
}
\email{rob@robjohnson.io}

\author{Prashant Pandey}
\affiliation{
\institution{Northeastern University}
  \country{USA}
}
\email{p.pandey@northeastern.edu}


\renewcommand{\shortauthors}{Bender et al.}


\begin{abstract}
Filters such as Bloom, quotient, and cuckoo filters are fundamental building blocks providing space-efficient approximate set membership testing. 
However, many applications need to associate small values with keys—functionality that filters do not provide. 
This mismatch forces complex workarounds that degrade performance.


We argue that \emph{maplets}---space-efficient data structures for approximate key-value mappings---are the right abstraction. 
A maplet provides the same space benefits as filters while natively supporting key-value associations with one-sided error guarantees.
Through detailed case studies of SplinterDB (LSM-based key-value store), Squeakr (\kmer counter), and Mantis (genomic sequence search), we identify the common patterns and demonstrate how a unified maplet abstraction can lead to simpler designs and better performance.
We conclude that applications benefit from defaulting to maplets rather than filters across domains including databases, computational biology, and networking.





\end{abstract}

\maketitle

\sloppy 













\section{Introduction}


A \defn{filter} provides an approximate representation of a set, and achieves space efficiency by allowing one-sided errors in membership queries. 
Filters, such as Bloom~\cite{Bloom1970}, quotient~\cite{Bender2012, Pandey2017}, cuckoo~\cite{Fan2014}, 
xor~\cite{Graf2020}, ribbon~\cite{Dillinger2022}, 
and many others, have become standard tools in systems builders' toolboxes.
In database systems, filters accelerate range queries in systems like SuRF~\cite{zhang2018}, reduce I/O costs in LSM-based storage engines like RocksDB and Cassandra~\cite{Cao2020}, enable efficient joins in distributed systems, and optimize query processing in modern analytical databases~\cite{diaconu2013}.
Filters have also found applications in networking~\cite{BroderMi04}, computational biology~\cite{Pandey2018, Pandey2017a, Pandey2017b}, and model checking~\cite{Dillinger2009}, among many others.

In this paper, we argue that filters are the wrong abstraction for many systems that use them.
The core problem is that filters represent \defn{sets}, but many applications actually need \defn{maps}, i.e. the ability to associate values with keys and retrieve them during queries.
This mismatch forces systems designers to employ complex workarounds to get the space savings of filters with the functionality of a map.
In addition to added complexity, these workarounds often limit functionality and degrade performance.

A \defn{maplet} is a space-efficient representation of a map $M$, analogous to how a filter is a space-efficient representation of a set.
When a maplet is queried with a key $k$, it returns a value $m[k]$ that is an approximation of the true value $M[k]$.
Specifically, $m[k] = M[k]$ with probability at least $1-\epsilon$.
Moreover, like a filter, a maplet allows for one-sided errors: it guarantees that $M[k] \preceq m[k]$ for some application-specific ordering relation $\prec$ on the map's values.
For instance, if the map stores counts, the maplet always returns a count at least as large as the true count.
If the values are sets, it returns a superset of the true set.


Maplets have appeared in various forms across different domains, often as domain-specific solutions to particular problems.
For example, database systems like SlimDB~\cite{Ren2017}, Chucky~\cite{Dayan2021}, and SplinterDB~\cite{Conway2020}, which we discuss in detail below, all employ data structures that associate values with keys while allowing approximate queries---essentially implementing maplets of varying sophistication.
However, these solutions have typically been developed independently for specific use cases
rather than being recognized as instances of a general, reusable abstraction.
\emph{The goal of this paper is to identify the common patterns across these applications and demonstrate how a unified maplet abstraction can lead to simpler designs and better performance across diverse domains.}



We demonstrate how maplets improve data-system design and performance through three positive examples where systems successfully adopted maplet-based approaches: (1) SplinterDB, an LSM-based key-value store that uses maplets to optimize query performance, (2) Squeakr, a \kmer counter for genomics data that achieves order-of-magnitude performance improvements with maplet-based counting, and (3) Mantis, a genomic search engine that leverages maplets to provide scalable sequence search across massive datasets.
These examples illustrate the practical benefits and design simplifications that maplets enable across diverse application domains.


To demonstrate the ubiquity of maplets, we also give sketches of how over half the networking applications of filters in Broder and Mitzenmacher's famous survey~\cite{BroderMi04} could be improved by using maplets instead.  
In fact, many of the applications are essentially cobbling together a poor-man's maplet from a collection of filters.

Our examples also demonstrate the importance of other properties of a full-featured maplet.
In particular, a full-featured maplet has (1)~value support, (2)~resizing, (3)~incremental construction, (4)~mergeability, (5)~deletion, (6)~space efficiency, and (7)~good cache locality.
Many of these features have been identified as useful features for filters.
As the examples in this paper will show, data systems need these features in maplets as well.



Before diving into the examples, we first give a simple maplet construction and, based on the construction, define a stronger version of the maplet error guarantee, which we will call a \defn{strong maplet}.

\section{How to Make a Maplet}
\label{sec:maplets}

We now describe a general technique for building a maplet from a \defn{perfect-hashing filter}, defined below.  
Numerous modern filters, including quotient~\cite{Pandey2017}, cuckoo~\cite{Fan2014}, and Morton~\cite{Breslow2018}, are perfect-hashing filters, so this technique is widely applicable.

\para{Perfect hashing.}  A \defn{perfect hash} $t$ is a data structure representing an injection from some set $\mathcal{T}$ to $\left\{0, \ldots, n\right\}$, where $n$ is at least as large as $|\mathcal{T}|$.  Since $t$ is injective, it can be used to map elements of $\mathcal{T}$ to slots in an array $A$ without collision. It is important for $n$ to be as close to $|\mathcal{T}|$ as possible, since larger $n$ would waste memory on unused slots of $A$.  Note that a perfect hash may return any value when queried on a key not in $\mathcal{T}$.

\para{Perfect-hashing filters and maplets.}
A \defn{perfect-hashing filter} is a filter that induces a perfect hash on the set of \defn{fingerprints} $h(\calS)=\left\{h(x)\mid x\in \calS\right\}$, where $h$ is some hash function and $\calS$ is the set of items currently in the filter.  Several filters meet this criteria implicitly. They maintain an array of size at least $|h(\calS)|$ slots, and queries search through the slots until they find a slot whose contents match the queried item.  The exact details of the search algorithms are not important here.  The important property is that, for each fingerprint in $h(\calS)$, there is exactly one slot that is a match.  Thus we can view the filter as assigning a fingerprint $h(k)$ to the index of the slot that it matches in the filter.  
For a perfect-hashing filter $F$, we will use $F[k]$ to denote the index that $F$ assigns to $h(k)$, and we will say that $F[k]=\perp$ whenever the filter indicates that $k$ is not in $\calS$.
Note that, although $F$ maps fingerprints to indexes injectively, it may not be an injection from \emph{keys} to indexes, since there may be collisions under $h$.


We now describe a general technique for building a maplet from a perfect-hashing filter, assuming that the values have some associative and commutative binary operator, $\oplus$, that respects the ordering of the values, i.e., $a, b \preceq a \oplus b$ for all values $a$ and $b$.  For example, if the values are counters, then $\oplus$ is just addition.  If the values are sets, then $\oplus$ is set union.

The core idea is to simply store an array $V$ of values and use the filter to map keys to slots of $V$, i.e., querying key $k$ in the maplet $m$ will return $m[k]=V[F[k]]$ whenever $F[k]\not=\perp$, and $m[k]$ will be $\perp$ otherwise.  Whenever we insert a kv-pair $(k,v)$, if $F[k]\not=\perp$, then we set $V[F[k]] \leftarrow V[F[k]] \oplus v$.  Otherwise we insert $k$ into the filter and then set $V[F[k]]\leftarrow v$.  Note that, if the values with $\oplus$ form a group, then we can also support deletion of a key-value pair $(k,v)$ by setting $V[F[k]]\leftarrow V[F[k]] \oplus -v$.

When the values do not form a group, then it may be impossible to implement deletion using the above scheme because, without a group structure, it may not be possible to cancel out some value $v$ that was previously merged into a slot of $V$.

Fortunately, most if not all perfect-hashing filters can also be extended to support \emph{multisets} of fingerprints, i.e., we can view $h(\calS)$ as a multiset in which a fingerprint may occur more than once due to hash collisions or because $\calS$ itself was a multiset.  In this case, a fingerprint that occurs $\ell$ times in $h(\calS)$ will be assigned to $\ell$ slots in the filter, but no slot will be assigned to two different fingerprints.

We can use this to implement a maplet that supports deletes as follows.

Whenever the user inserts a new key-value pair $(k, v)$, we simply add $h(k)$ to the filter.  This will allocate a new slot of $V$ to the fingerprint $h(k)$.  We store $v$ in that slot.  For a query of key $k$, we find all the values $v_1, \ldots, v_\ell$ associated with $h(k)$ and return $v_1 \oplus \cdots \oplus v_\ell$.  Deletion of a key-value pair $(k, v)$ can be implemented by simply removing one instance of $h(k)$ from the filter and rearranging the values in $V$ so that one copy of $v$ gets deallocated from $h(k)$.

\para{Space and error rates.}  The total space for such a maplet will be the same as the space for the underlying filter, plus the space for $V$.  If we assume each value can be encoded in $v$ bits, then the per-item space using a quotient or cuckoo filter will be $O(\log 1/\epsilon + v)$ bits, i.e., there will be $v$ bits of overhead to store the value.

The error rate will be at most $\epsilon$.  
To see why, consider a perfect-hashing maplet $m$ representing map $M$.
Then, for any $k$, $m[k]=M[k]$ unless there exists a $k'\not=k$ in $M$ such that $h(k')=h(k)$.
But for any given  $k'$ in $M$, the probability that $h(k')=h(k)$ is $\epsilon/n$ so by the union bound the probability that there exists a  $k'$ in $M$, where $h(k')=h(k)$ is at most $\epsilon$.

\para{Practical considerations.}  In practice, we don't need a separate vector $V$.  Rather, most if not all perfect-hashing filters maintain some array internally.  We can just expand the entries in those arrays to also store the values associated with the fingerprints that map to those slots.  This has the advantage of better cache locality, and it means that, as the filter shuffles around entries in its array, the values will correctly move along with them.

This scheme also supports all the same operations as the underlying filter.  If the filter supports deletions, then so does the maplet.  If the filter supports merging or resizing, then so does the maplet.  The maplet also inherits the filter's data locality, 
enumerability, and all other properties.  Thus, for example, by applying this technique to a quotient filter~\cite{Pandey2017}, we can get a full-featured maplet that meets all the requirements outlined in the introduction.
It further supports efficient counting using a variable-length encoding of counters (see~\cite{Pandey2017} for a detailed explanation). 

We note that a problem arises when building a multiset maplet from a cuckoo filter.  The cuckoo filter supports at most 8 copies of a given fingerprint, and hence we can insert at most 8 values associated with a single fingerprint. This can be a real problem for some applications that wish to support deletions but for which the $\oplus$ does not form a group.

\para{Strong maplets.}
The above constructions actually offer a stronger error guarantee than simply returning the correct value with probability at least $1-\epsilon$.  
In addition to usually returning the right answer, they also return answers that are ``close to right'' with high probability.

More formally, they guarantee what we call the \defn{strong maplet property}, i.e., that 
$$m[k]=M[k]\oplus\left(\bigoplus_{i=1}^{\ell}M[k_i]\right)$$
for some set of $\ell$ keys $k_1,\ldots, k_\ell$ in $M$, where $\Pr[\ell\geq L] \leq \epsilon^L$.  In other words, they guarantee, with high probability, that the result of a query is the result of merging a small number of values in the map.  So, for example, if all the values in $M$ are singleton sets, then this guarantees that the probability of having $\ell$ extra items in the set returned by a query to $m$ falls off exponentially in $\ell$.  In other words, even when the maplet is wrong, it isn't very wrong.

\section{Storage Engines \& Databases}

In this section we show how maplets can be used to break through prior tradeoffs to achieve better storage-engine performance.

Storage engines are the component of database management systems (DBMS) which manage the underlying data.
They often represent the primary performance bottleneck, making their optimization crucial for overall performance. 
As a result, different storage engines employ distinct architectures optimized for specific workload characteristics.

Log-structured merge-trees (LSMs)~\cite{ONeil1996} and their variants, including $B^\varepsilon$--trees~\cite{Brodal2003lower}, have become a popular storage engine architecture for write-intensive applications. 
Major systems including RocksDB, Cassandra, HBase, and SplinterDB~\cite{Conway2020, Conway2023} all employ LSM-based designs.

LSMs are designed to optimize write throughput.
Rather than immediately modifying data on disk, incoming writes are first accumulated in a fast, in-memory structure called a \defn{memtable}.
When the memtable reaches capacity, its sorted contents are flushed to disk as an immutable \defn{sorted string table} (SSTable).
Read operations check the current memtable first, then search through SSTables in reverse chronological order. 
A background compaction process continuously merges and reorganizes SSTables to maintain read performance and reclaim space, ensuring the system remains efficient as data accumulates over time.
This design enables all disk writes to use large sequential I/O, optimizing for update performance.


LSM-based storage engines face a fundamental trade-off between write and read performance, determined by their compaction policy~\cite{CompactionPolicies}.
This trade-off is characterized by two key metrics: \defn{write amplification} (the total number of times data is rewritten during its lifetime) and \defn{read multiplicity} (the number of SSTables that must be checked to locate a key).
Different compaction policies balance these costs in distinct ways~\cite{CompactionPolicies}.

The two dominant compaction strategies are \defn{leveled}~\cite{ONeil1996} and \defn{size-tiered}~\cite{DBLP:conf/vldb/JagadishNSSK97} policies, both organizing the LSM into $h$ levels $L_i$ with geometrically growing capacity ($|L_{i+1}| = g|L_{i}|$).
Leveled compaction maintains non-overlapping SSTables within each level, which requires items to be compacted up to $g$ times per level (yielding $gh = g\log_g n$ write amplification) but requires reading only one SSTable per level ($h = \log_g n$ read multiplicity).
Size-tiered compaction flushes and compacts entire levels together, compacting each item only once per level ($h = \log_g n$ write amplification) but allowing up to $g$ overlapping SSTables per level, which means that queries may have to read up to $g$ SSTables per level ($g\log_g n$ read multiplicity).
Thus, leveled LSMs optimize for read performance while size-tiered LSMs optimize for write performance.

\mab{Whereas a filter can say yes/no given a particular location, a maplet can just tell you the key's location is, given a single query.}

\mab{In discussion with Prashant and Rob: \\
Value support-- associate value which points to all SST tables where a key is (might be) present. \\
Good cache-locality: many situations with memory pressure, even filters are paged out to disk.
We don't need resizing or deletes.
May not even need incremental construction.}

\para{Employing filters in storage engines.}
Nearly all LSM-based storage engines employ filters as a critical optimization, maintaining a filter for each SSTable to avoid expensive storage reads during queries.
This optimization proves especially valuable in configurations with high read multiplicity, such as size-tiered LSMs, where queries must potentially check multiple SSTables per level. 

However, modern NVMe storage has dramatically reduced random-access latency, fundamentally changing LSM query cost profiles.
While queries now require only 0-1 storage reads due to effective filtering, they still need 5-40 filter queries depending on read multiplicity~\cite{Conway2023}.
As a result, the total cost of querying filters can be one of the main bottlenecks to query latency, particularly in size-tiered LSMs with high read multiplicity.

This problem has inspired researchers to develop some early examples of maplet-like data structures.
SlimDB~\cite{Ren2017} uses per-level cuckoo filters that store SSTable identifiers as values, reducing filter queries to one per level.
However, SlimDB adds complex compaction logic and multi-level filters to handle duplicate filter entries, in order to limit tail latency as well as to sidestep cuckoo filters' limited key-multiplicity support.
This compaction logic can impact overall write performance under some workloads.
Chucky~\cite{Dayan2021} employs a global cuckoo-filter-based mapping of keys to SSTables, enabling a single filter query to locate the SSTable that must be searched for a given key.
However, this approach requires constant filter updates during compaction, adding significant system complexity.
Furthermore, these updates are to random locations in the filter, which means the filter must fit in RAM to get good performance.
Consequently, Chucky employs sophisticated value compression techniques in order to squeeze every last bit out of the filter.

\para{Memory pressure.}
Despite being much smaller than the underlying data, filters on large datasets still impose significant memory pressure. Extensive research has sought to optimize the trade-offs between false positive rates, latency, and memory usage~\cite{Pandey2017, Pandey2021, Dayan2017}, but fundamental theoretical lower bounds dictate that filters require at least 8-12 bits per item~\cite{doi:10.1137/120867044, Pandey2021} to achieve typical false-positive rates.
In practice, both large datasets and relatively small key-value pairs are common~\cite{Cao2020}, so this translates to a total filter size of 1--2\% of the dataset size, demanding expensive high-memory servers and consuming RAM that could be used for other purposes. 

When filters cannot fit in memory, their utility disappears since accessing a filter on storage costs as much as directly querying the underlying SSTable.
However, systems like Chucky and SlimDB justify this memory requirement because their filters can eliminate multiple SSTable accesses per query.
Nevertheless, both systems mandate in-memory filter storage because disk-resident filters would incur prohibitive update costs during compaction operations, particularly when resolving key multiplicity issues.

\mab{I don't love the following title of the section heading as maplets. Maybe the ways in which maplets help or something else..}

\para{How maplets help storage engines.}
SplinterDB builds on the line of research started with SlimDB and Chucky by incorporating maplets into a size-tiered LSM instead of filters~\cite{Conway2023}.
In SplinterDB, each level is broken up into nodes, and each node contains a maplet indicating in which SSTable each key in the node can be found.
Each maplet replaces a collection of filters, and setting the false-positive rate for the maplet to the collective false-positive rate of those filters results in the same memory usage---essentially the bits that were used for FPR reduction are exactly used for the SSTable ids~\cite{Conway2023}.

Importantly, maplets can be paged out to storage, and then only the blocks required to answer a query can be paged back in at query time.
This typically incurs a single IO, which can save the up to $g$ IOs required to query the SSTables.
Therefore, maplets improve query performance even when paged out to storage.
When a new SSTable is flushed into the node as part of a compaction process, the maplet can be updated with the new keys without accessing any present data from other SSTables.
As a result, only the maplet needs to be read from storage during the update process.
Maplets can be updated using sequential access, so that the entire maplet need not be in memory during updates~\cite{Conway2023}.

\section{\kmer counting}


Numerous analyses in computational biology begin by counting the number of occurrences of each \kmer (length-$k$ subsequence) in a sequencing dataset,
in order to weed out erroneous data caused by sequencing errors.
These errors typically produce \defn{singleton} \kmers that appear only once in the dataset, and since the number of singletons grows linearly with dataset size, \kmer counting becomes essential for identifying and filtering out these artifacts.
\kmer counting is further employed for estimating sequencing depth and preprocessing data for genome assembly.

\kmer counting presents significant computational challenges due to the combination of large dataset sizes, the need for fast processing, and highly skewed \kmer frequency distributions.
These constraints have led to diverse proposals for system architectures that optimize for different performance objectives, including minimizing memory consumption, improving cache locality, and enabling efficient parallel processing across multiple threads.
These approaches have been used in popular \kmer-counting tools such as 
Jellyfish~\cite{MarccaisKi11}, BFCounter~\cite{Melsted11Efficient}, and Turtle~\cite{RoyBhSc14}.
%

Several systems
use a Bloom filter to identify and filter out singleton \kmers, thus reducing the memory consumption~\cite{MarccaisKi11,Melsted11Efficient,RoyBhSc14}
Then they use larger hash tables, or other traditional data structures, for the actual counting.
Thus, \kmers are inserted into the Bloom filter upon first observation, and then stored in a hash table along with their counts upon subsequent observations.

Bloom filters impose significant limitations on \kmer-counting systems.
First, they only support membership queries and cannot count keys, forcing systems to maintain a separate data structure for counting.
This dual-structure approach requires every insertion to access two data structures, degrading performance.
Second, Bloom filters suffer from poor cache locality, leading to suboptimal memory access patterns during high-throughput \kmer processing.
Third, Bloom filters lack support for essential operations like resizing and deletion, forcing applications to over-provision memory to accommodate worst-case scenarios, which sacrifices the space efficiency that made Bloom filters attractive in the first place.
As we articulate below, an integrated counting data structure, using a maplet, often proves more space-efficient than the Bloom filter plus hash table combination that these systems are forced to employ.

The Bloom filter has inspired numerous variants that try to overcome one drawback or the other~\cite{QiaoLiCh14, PutzeSaSi07, LuDeDu11, DebnathSeLi11, AlmeidaBaPr07}. For example, the counting Bloom filter (CBF)~\cite{BonomiMiPa06} replaces each bit in the Bloom filter with a $c$-bit saturating counter, allowing it to support deletes, but increasing the space by a factor of $c$.
However, none of these variants overcome all the drawbacks.


\para{How maplets help \kmer counting.}
Squeakr's~\cite{Pandey2017b} maplet-based approach using the Counting Quotient Filter (CQF)~\cite{Pandey2017} provides a unified solution.
The CQF maplet natively associates count values with \kmer keys, eliminating the need for separate membership and counting structures and simplifying the system architecture.

The performance benefits are substantial: Squeakr achieves order-of-magnitude faster queries and up to $4\times$ faster construction compared to state-of-the-art \kmer counters, while producing smaller indexes.
These improvements stem from several maplet advantages: (1)~value support---by storing counts efficiently, (2)~avoiding the memory over-provisioning required with fixed-size Bloom filters, (3)~supporting essential operations like incremental construction and deletion that Bloom filters lack, (4)~providing better cache locality through integrated storage.

Additionally, the CQF maplet's support for thread-safe counting and efficient scalability across multiple threads---even with highly skewed \kmer distributions---enables Squeakr to fully utilize modern parallel hardware, which is difficult to achieve with the dual Bloom filter plus hash table approach.

\section{Large-scale sequence search}

The Sequence Read Archive (SRA)~\cite{LeinonenSuSh10} represents one of the largest publicly-available biological datasets, containing over 50 petabases of raw RNA-Seq data from millions of experiments worldwide. This vast repository holds the potential to revolutionize biological discovery by enabling researchers to search for evidence of newly-discovered genes, identify novel viral sequences, investigate disease-associated mutations, and validate experimental findings across the entire corpus of publicly-deposited sequencing data.

To unlock this potential, researchers have developed specialized tools for large-scale sequence search across the SRA~\cite{Solomon2016, HarrisM20, Pandey2018}. These data systems enable scientists to query the entire SRA for occurrences of specific transcripts or gene sequences, transforming what was once an archive of isolated experiments into a queryable resource for biological discovery. The ability to search across millions of experiments allows for rapid validation of gene function hypotheses and accelerated identification of biological patterns that would be impossible to detect within individual datasets.

Solomon and Kingsford~\cite{Solomon2016} formalized the \emph{experiment discovery problem}: given a transcript sequence of interest, decompose it into constituent \kmers to form a query set $q$, then identify all SRA experiments that contain at least a user-specified fraction $\Theta$ of these query \kmers. This problem captures the common biological workflow of searching for experimental evidence supporting a particular gene or transcript.

To solve this problem efficiently, they introduced the sequence Bloom tree (SBT)~\cite{Solomon2016, HarrisM20}, a hierarchical index structure. The SBT organizes experiments as leaves in a binary tree, where each leaf contains a Bloom filter representing the \kmers present in that experiment. Interior nodes store Bloom filters representing the union of all \kmers in their subtrees, enabling efficient pruning during search. However, because Bloom filters produce false positives, the SBT inherits these errors, leading to false positive experiments in the final search results.

Subsequent research has proposed numerous improvements to the original SBT~\cite{SolomonK17, sun2017allsome, bradley2019ultrafast, yu2018seqothello, gupta2021fast, bingmann2019cobs}, all attempting to optimize the Bloom filter representation for better space efficiency and query performance. The primary limitation 
\mab{not quite the right word...  This is one place where the system designer is tied themselves into pretzels to work around the limitations of the Bloom filter}
of the original SBT was its inefficient use of space due to uniform-sized Bloom filters throughout the tree structure. SSBT~\cite{SolomonK17} addressed this by ``splitting'' the Bloom filters into two distinct filters that store unique subsets, along with introducing ``noninformative bits'' that can be pruned when they provide no discriminatory value. HowDe-SBT~\cite{HarrisM20} further refined this approach by ``minimizing the number of active positions and 1 bits to improve space'' through more sophisticated bit vector representations.
\mab{that is, because they didn't know how many elements they were putting into the bloom filter, they had to overprovision, leading to a bloom filter that was space and efficient, and mostly had zeros, so then they compress the Bloom filter, so that it didn't waste space.}

However, all these approaches face fundamental limitations inherent to Bloom filters. The tree structure amplifies false-positive rates toward the root, where filters must represent the union of many experiments. Additionally, since all Bloom filters must maintain uniform size to support merging operations, filters at leaf nodes representing large experiments suffer from high false-positive rates that distort query traversal paths.


\para{How maplets help sequence search.}
\mab{in this example, maplets are used to store subsets of a large set of experiments. We also use amny of the other features of a full-featured maplet to gain performance.}
Maplets enable Mantis~\cite{Pandey2018} to eliminate the 
fundamental 
architectural complexity that plagues SBT-based approaches.
Mantis uses a maplet-based inverted index that directly maps each \kmer to its associated experiments. 
The counting quotient filter (CQF) maplet associates each \kmer with a bit vector indicating exactly which experiments contain that \kmer, creating a 
straightforward one-to-one mapping that eliminates the need for tree traversal and threshold tuning.

This maplet-based design provides exact results rather than approximate ones. Unlike SBT systems that suffer from accumulating false positives as they traverse up the tree structure, (maplet-based) Mantis employs fingerprints that match the original key size, enabling the maplet to function as an exact map. This eliminates the uncertainty inherent in SBT results where users must interpret approximate hit ratios and adjust thresholds to avoid false discoveries.

The performance benefits are substantial: the maplet-based system, Mantis, proves to be smaller, faster, and exact compared to SBT-based approaches. The maplet's direct key-to-value mapping eliminates the computational overhead of tree traversal, bit operations across multiple levels, and the complex pruning algorithms required by SBT variants. Additionally, the subsequent versions of Mantis successfully indexed up to 40K experiments from the SRA comprising over 100TB of sequencing data, demonstrating that the maplet approach scales efficiently to massive datasets while maintaining exact results and fast query performance.

\mab{We need value support in order to associate downstream the list of experiments where a \kmer appears.
We need resizing because we don't know a priori how many \kmers there are. 
And we did incremental construction because we don't a priori how many \kmers there are.
We need a good cache locality to quickly query the index to retrieve all the relevant experiments given a query sequence. 
We don't need deletion in Mantis, but we do need deletion in Sqeakr to wead out \kmers based on count. }

\mab{Just to be explicit about the thing, all these comments are in discussion with Prashant, and mostly from him.}

\section{Networking Applications}
\label{sec:networking}

We demonstrate maplets' broad applicability by showing how they can improve over half the networking applications in Broder and Mitzenmacher's classic Bloom filter survey~\cite{BroderMi04}. These improvements stem not from using newer filter variants, but from fundamentally restructuring applications around the maplet abstraction.


Broder and Mitzenmacher begin by describing how Bloom filters are used in Fan, et al.'s Summary Cache~\cite{FanCaAl00}. 
The Summary Cache system~\cite{FanCaAl00} illustrates typical Bloom filter usage: web caching proxies exchange Bloom filters representing their cache contents, and on cache misses, query each peer's filter to find potential sources. This requires maintaining counting Bloom filters locally (to handle cache additions/deletions) and constructing regular Bloom filters for transmission, with delta encoding to reduce space~\cite{BroderMi04}. False positives simply result in rejected requests or redundant web fetches.


Maplets solve two key problems in this scheme. First, instead of querying each peer's filter individually (linear in the number of peers), a single maplet can map keys to the set of peers that have them, enabling one-query peer discovery. Second, rather than maintaining space-inefficient counting Bloom filters and constructing transmission filters, proxies can maintain a maplet of cache \textit{changes} since the last update, mapping key hashes to add/delete counts. This delta-based approach eliminates the space overhead of full cache filters, avoids filter construction costs, and provides delta encoding benefits automatically.



The above example is emblematic of a whole class of resource routing applications in Broder and Mitzenmacher's survey. In section 5.1, Broder and Mitzenmacher describe the use of filters to record the locations of every data item stored in a peer-to-peer network (i.e., every peer keeps a Bloom filter of the keys of all items stored at every other peer).  This has the same problem as above: to find an item, you have to query a lot of Bloom filters.  Maintaining a maplet mapping each key to the peers that have that key would enable much faster lookups.

Section 6.1 describes a variation in which the nodes in the network are organized into a rooted tree.  Each node $x$ maintains a filter of the contents of each of its subtrees and sends its parent a filter representing the set of all items stored in the subtree rooted at $x$. 
Nodes use their unified list to check whether a requested item is in their subtree. If not, they route the request to their parent.  Otherwise, they use each of their children's filters to determine which child to forward the request to.  Section 6.3 describes a special case of this general idea in which the tree is a quad tree.

Again, maplets solve multiple problems. Each node could maintain a maplet mapping keys in its subtree to the child containing them.  Now a single maplet query can replace the task of querying each child's filter.  Furthermore, the filter-based approach has the same space problems as the Sequence Bloom Tree~\cite{Solomon2016}: each node must OR all its children's filters together, which means that all the Bloom filters in the network must be the same size, even though filters at the leaves will have far fewer items than filters near the root.  This means that filters will either be too sparse, be too dense, or require extra steps, such as compression, to achieve space efficiency. Resizable, mergeable maplets solve this problem automatically.

In Section 6.2, Broder and Mitzenmacher summarize the attenuated Bloom filter, which is used in OceanStore~\cite{KubiatowiczBiCh00} to find the shortest route to a given requested item.  It involves each host keeping an array of filters associated with each of its outgoing edges. To find a route, a host must query all the filters in all the arrays until the item is found.  A maplet mapping each key to a set of ordered pairs (edge, distance) would enable routing with a single maplet query.

Section 7.3 describes a technique for implementing multicast in which a router maintains, for each outgoing edge, a Bloom filter of all the multicast addresses that should be forwarded along that edge.  The same information could easily be encoded in a maplet, so routing could be done with a single maplet query. 
We do note that, in a hardware-based router implementation, all the Bloom filter checks can be performed in parallel, and so filters may be the better choice in a hardware implementation. Nonetheless, a maplet could be substantially faster in a software router.

Section 8.2 describes an IP traceback mechanism in which each router keeps a Bloom filter of packets it has recently forwarded.  The traceback algorithm queries every neighbor of a router to determine which ones might be the next hop towards the packet's source.  The scheme could be improved by having a maplet that maps packet IDs to the interface on which they arrived.  This might appear to require more space than simply keeping a Bloom filter of observed packets but, as in the application in SplinterDB~\cite{Conway2020}, keeping the origin information means we can use a higher false-positive rate, so that the total space is the same.  The branching factor due to false positives will be exactly the same as with the filter-based approach, but the advantage is that, during traceback, we do not need to query every neighbor of a router, rather we need to query only the neighbors listed for that packet in the maplet.


\bibliographystyle{ACM-Reference-Format}
\bibliography{bibliography}


\end{document}